\documentclass{article}
\usepackage{amsmath}
\usepackage{amsfonts}

\begin{document}

\noindent Publsihed in \textit {Philosophy of Science} \textbf{80}, 22-48 (2013).\footnote{%
To contact the author, write to Merton College, Oxford OX1 4JD, or email simon.saunders@merton.ox.ac.uk. My thanks for encouragement, corrections, and criticisms to Julian
Barbour, Harvey Brown, Oliver Pooley, Robert DiSalle, George Smith, and
David Wallace. I owe a particular debt of gratitude to Julian: in
conversations and writings he has long influenced my thinking on space-time
matters. }\bigskip
\bigskip
\begin{center}
{\LARGE Rethinking Newton's \textit{Principia}\bigskip }

{\Large Simon Saunders}
\bigskip
\end{center}

\begin{abstract}It is widely accepted that the notion of an inertial frame is central to Newtonian mechanics and that the correct space-time structure underlying Newton’s methods in Principia is neo-Newtonian or Galilean space-time. I argue to the contrary that inertial frames are not needed in Newton’s theory of motion, and that the right space-time structure for Newton’s Principia requires the notion of parallelism of spatial directions at different times and nothing more.
\end{abstract}
\bigskip
{\Large 1. Introduction\smallskip \bigskip }

\noindent Few writings in physics have occasioned as much philosophical
interest as Newton's \textit{Principia, }but still puzzles remain. One is an
alleged inconsistency in Newton's stated principles in application to a
universe that is, quite possibly like the actual universe, infinite and
approximately homogeneous in space.\footnote{%
See e.g. Norton [1993], Vickers [2009]. For the early history of this
problem, see Norton [1999].} Another is that the motions identified by
Newton's methods as privileged (`true', according to Newton) could not,
realistically, be considered even as motions relative to an inertial frame%
\footnote{%
This question was first raised by Kant. I shall come back to Kant's views
shortly.} -- or not if the universe is sufficiently large.

The two problems are evidently related. Indeed, if the material universe is
in fact infinite and if Newton's theory really is inconsistent in that
application\ we should expect that\textit{\ no} true motions in Newton's
sense could have been identified. But that raises a further puzzle in its
own right. For Newton surely did pick out definite motions as somehow
preferred; his theory as laid down in the \textit{Principia} (what I shall
call Newton's theory of motion or NTM) was empirically successful; it was
genuinely informative: How can an inconsistent theory be empirically
informative?\ 

Of course there are other cases of alleged inconsistencies in empirically
successful theories. Relativistic quantum electrodynamics, with its famous
divergences,\ is another example. It is tempting to place NTM in the same
category, and to view the inconsistency problem as an instance of a more
general puzzle (of how inconsistencies can in practise be `contained'). That
has been the tenor of several recent discussions of the subject.\footnote{%
An exception is Malament [1995], who showed how the problem can be solved
(effectively by going over to Newton-Cartan theory; see below).}

That is not the strategy taken here. My objective is to give a consistent
reading of Newton's \textit{Principia -- }but\textit{\ }largely discarding
the Scholium to the Definitions -- under which it applies equally to an
infinite as to a finite mass distribution, and hence (if only in the
non-relativistic limit) to the actual universe. On this reading the nearest
thing to an inertial frame is a non-rotating frame with arbitrary (and
possibly time-dependent) linear acceleration. The privileged frame (the
centre of mass of the solar system), wrongly identified by Newton as
inertial, indeed as at rest, is exemplary: it is a local freely-falling
non-rotating frame. The latter, moreover, is just the sort of frame that is
privileged in Einstein's general theory of relativity (GTR),\footnote{%
See Knox [2013, 2014] for a defence of the claim that such frames play the same
functional role in GTR that inertial frames in practise played in NTG.} and
in Cartan's non-relativistic theory of gravity (so-called Newton-Cartan
theory or NCT). These theories have more in common with NTM than is normally
thought.

There is more. As John Norton has recently explained (Norton [1995]),
acceleration in NCT in an infinite universe is \textit{relative}, not
absolute. On the reading of \textit{Principia} I shall propose the same is
true, but in a rather more explicit sense. It is that accelerations, as all
other motions, are relational in roughly the sense of the
absolute-relational debate in Newton's time.\footnote{%
Although it is not motivated in the same way.} This reading extends to any
theory satisfying Newton's laws, yielding a form of relationalism that
applies to the entire scope of classical particle mechanics and Newtonian
gravity.\footnote{%
I also believe that it extends to non-relativistic field theory, but I will
not try to defend that claim here. See Hood [1970], Rosen [1972] for the
construction of such a theory, specifically in non-relativistic quantum
mechanics.} In this respect it is unlike relationalism in the Machian sense,
as developed by Julian Barbour.\footnote{%
See Barbour [1999] for a masterly and self-contained introduction.} The
latter has the following feature:\ only systems which (in NTM terms)\ have
total energy and angular momentum zero can be defined. No such constraint --
and no such prediction -- follows from relationalism in our sense.

What about relationalism in Norton's sense, in NCT? To explain this we will
need to use the modern apparatus of differentiable geometry, and the concept
of a connection $\Gamma $ -- a rule for defining differentiation of vectors
by continuous transport along smooth curves. Thus, suppose there is a
manifold $M$ equipped with global rectilinear coordinates $x^{\alpha
}=\langle x^{0},x^{j}\rangle ,$ $j=1,2,3$ (so $M$ has the topology of a
Euclidean space). The coordinates $x^{\alpha }$ are eventually to be
identified as inertial coordinates. In terms of these, let the connection $%
\Gamma $ have as its only non-zero components the quantities: 
\begin{equation}
\Gamma _{00}^{j}=\frac{\partial \phi }{\partial x^{j}}
\end{equation}%
where $\phi $ (the gravitational potential) satisfies: 
\begin{equation}
\nabla ^{2}\phi =-4\pi G\rho
\end{equation}%
with $G$ is the universal gravitational constant $G=6.67\times 10^{-11}$ $%
m^{3}\sec ^{-2}kg^{-1}$ and $\rho $ the mass density. The equation of motion
for a point particle with curve $x:\mathbb{R}\rightarrow M$, parameterized
by real numbers $\lambda $, is then the same as in GTR; it is the geodesic
equation: 
\begin{equation}
\frac{d^{2}x^{\alpha }}{d\lambda ^{2}}=\Gamma _{\beta \gamma }^{\alpha }%
\frac{dx^{\beta }}{d\lambda }\frac{dx^{\gamma }}{d\lambda }.
\end{equation}%
Affine geodesics are curves whose tangent vector $\frac{dx^{\beta }}{%
d\lambda }$ at each point is the parallel transport of the tangent vector at
any other point. Curves that depart from these represent the motions of
accelerated bodies.

One can free the theory from dependence on a special coordinate system by
introducing spatial and temporal metrics, satisfying suitable compatability
conditions, and by introducing a Riemann tensor as defined by the
connection. The theory can then be written in a generally covariant form as
in GTR.

The result is Newton-Cartan theory. It is more general than the theory based
on (1) through (3). When the total mass is finite (an `island universe') the
connection has a unique decomposition into a gravitational and inertial part:%
\begin{equation}
\Gamma =\Gamma _{grav}+\Gamma _{inertial}
\end{equation}%
such that there exist global rectilinear coordinates in which $\Gamma
_{grav} $ satisfies (1) and $\Gamma _{inertial}$ vanishes. Acceleration, as
defined by departures from straight-line motions with respect to $\Gamma
_{inertial}$ , is then uniquely defined. More prosaically, in the
coordinates in which $\Gamma _{inertial}$ vanishes, non-accelerating motions
are given by linear equations in $x^{0}$ and $x^{j}$. In this context (and
in this sense)\ acceleration is absolute. But $\Gamma $ can still be defined
even in the case of an infinite and homogeneous mass distribution. In that
case the decomposition (4)\ is no longer unique, there are no privileged
inertial frames, and there is no absolute meaning to acceleration.
Acceleration is only defined relative to a decomposition of the connection
into an inertial and a gravitational part, and in the case of an infinite
mass distribution, no such decomposition is preferred.

When I\ say NTM can be read as a relational theory, I mean something rather
different. In the first instance we are speaking of a particle theory, not a
field theory. The mathematics is also elementary -- it is a reading of 
\textit{Principia}. And my claim is that motions are relative whether or not
the matter distribution is infinite, in roughly the way relationalists like
Huygens and Leibniz thought it was. In the case of a finite number of
particles, it is true, the system as a whole can be taken to define a frame
of reference that is inertial as in the usual reading of \textit{Principia}.
Kant indeed took this as the only true inertial frame obtainable by Newton's
methods, with the solar system only the first step (yielding only a first
approximation) in an iterative procedure that in principle can only be
completed when the entire universe is taken into account. Mach, and arguably
Leibniz, thought the same. I\ am taking the opposite strategy to Kant's. I
am suggesting a reading in which this limit is not necessary to the
structure of the theory and plays no role in its application, and in which
the concept of inertial motion in the usual sense may not even be defined.

As emphasized by Michael Friedman [1992] and Robert DiSalle [2006], the key
to applying Newton's theory of gravity to only a part of the actual matter
distribution -- to beginning the iterative procedure envisaged by Kant as
concluding only with the entire universe -- is Corollary VI to the Laws. On
my reading of \textit{Principia} it is put to wider use:\ Corollary VI frees the
theory from the need to give any operational significance to the notion of
inertial frame. And it poses a further question. Newton thought absolute
space and time and absolute velocities were needed to make sense of his
Definitions and Laws, but the latter admitted more symmetries; they included
boosts among inertial frames as proved in Corollary V (the relativity principle).
As a result Newton's absolute space and time and absolute velocities were
dispensed with, and replaced by Galilean space-time and relative velocities,
with accelerations as the only absolute quantities. But the Definitions and
Laws, as applied to the motions of bodies `among themselves', have still
another symmetry, namely the group of time-dependent boosts among
non-rotating frames, as demonstrated by Corollary VI. What should then replace
Galilean space-time and absolute accelerations?\bigskip 

\noindent I begin with the passage from Newtonian to Galilean
space-time. This was consistent with all the propositions of Bk.1 and Bk.2
of \textit{Principia}, but it did violence to the Scholium to the
Definitions; it is a conservative extension of NTM all the same. It is our
template for what follows. In \S 3 I shall give an argument from cosmology
-- a near neighbour to the one I began with above -- to show we cannot in
practise determine any inertial motions, not in an infinite nor in a finite
universe (so long as it is sufficiently large), by Newton's principles -- so
knowing those motions cannot possibly have played any role in the actual
application of NTM to planetary motions, or to any others that have actually
been observed. The section following is on Corollary VI, and the explanation of
the irrelevance of inertial motions to Newton's model of the solar system
even in the favourable case of an island universe. In \S 6 and 7 the ladder
is kicked away. The last sections are on relationalism and the definition of
the `right' space-time for\textit{\ Principia}.\bigskip \bigskip 

{\Large 2. Newton's absolute space\bigskip \smallskip }

\noindent Newton worked hard, in the Scholium to the Definitions, to explain
why a theory of motion required an absolute, immaterial backdrop -- absolute%
\textit{\ }space -- that was not subject to motions of its parts or to any
change. He saw plainly the inadequacy of an appeal to the stellar background
or any other observable body as an absolute standard of rest: that could
define only a `relative space', a `movable dimension or measure' of motion
-- quantities that may have nothing to do with true motions. No observable
body was immovable. No idealization was safe.

The avowed aim of \textit{Principia}, as stated in the Scholium to the
Definitions, was to show how these true motions were to be determined,
including absolute velocities. The Laws that immediately followed spoke of
straight-line motions, and motions deviating from these -- absolute
accelerations. Non-accelerating motions were `straight' by reference to the
sequence of positions of bodies in absolute space at equal times, with equal
distances traversed in equal (absolute) times (in $4$-dimensional terms,
straight lines in Newtonian space-time). These locations in space, and space
itself, were part of the furniture of the universe even if they were
insensible. The relative motions of bodies, subject to forces satisfying the
Laws, were supposed to eventually yield the true motions. But by wide assent
what Newton's methods actually gave was motions referred to Galilean rather
than Newtonian space-time, in which no velocity is preferred.

Relationalists like Huygens and Leibniz went further. According to Leibniz,
space was `nothing at all without bodies, but the possibility of placing
them'; and `instants, consider'd without the things, are nothing at all; ...
they consist only in the successive order of things' (Alexander [1956
p.26-27]). His most damaging criticism was that according to Newton's
principles, supposedly determinate physical properties and relations were
undetermined by any possible measurement, however indirect -- not just
absolute positions but the velocities of particles too. They could be made
whatever you like -- there could be no `sufficient reason' as to why they
are thus and so -- nor could you ever know that they are thus and so.
Neither the speed nor the direction of the motion of the material universe
as a whole, with respect to absolute space, could have any empirical
meaning; hence neither could the absolute speed or absolute direction of
motion of any body within it.

The critique applies equally to Galilean (or for that matter to Minkowski)
space-time, if taken in the substantivalist sense (I shall come back to this
sense in a moment), but it is particularly telling in the context of
Newton's absolute space and time. There the problem can be stated
independent of Leibniz's principles and of substantivalism. In space-time
terms, Newton's absolute space picks out a preferred velocity -- absolute
rest -- but the laws of motion do not. The symmetry group of Newtonian
space-time is thus smaller than the symmetry group of the laws. The theory
itself says there are these absolute quantities, whose values can never be
obtained, for from the laws it follows (the principle of relativity):%
\footnote{%
I am taking the laws to include the requirement that mass is a scalar and
that force-functions do not depend on absolute positions or velocities, for
only then is Corollary V a corollary (see Barbour [1989 p.31-2], Brown [2005
p.37-8]). From sec. 5 on I shall, for simplicity, assume force functions do
not depend on relative velocities either.}\ 

\begin{center}
COROLLARY V
\end{center}

\begin{quotation}
\noindent The motions of bodies included in a given space are the same among
themselves, whether that space is at rest, or moves uniformly forwards in a
right line without any circular motion
\end{quotation}

\noindent Given that the motions of bodies among themselves -- if necessary,
all bodies -- are all that any possible observation can ever have to go on,
absolute velocities can never be determined.

Newton was clearly aware of this difficulty. For textual evidence,
consider:\ the third edition of \textit{Principia} contains only three
hypotheses, one in Book 2, two in Book 3, the first and third of a technical
nature. The second reads

\begin{center}
HYPOTHESIS I
\end{center}

\begin{quotation}
That the centre of the system of the world is immovable
\end{quotation}

\noindent It was followed with the remark:

\begin{quotation}
\noindent This is acknowledged by all, while some contend that the earth,
others that the sun, is fixed in that centre. Let us see what may from hence
follow.
\end{quotation}

\noindent The proposition which followed and its proof resolved Leibniz's
under-determination problem -- in a manner of speaking:\smallskip

\begin{center}
PROPOSITION XI. THEOREM XI
\end{center}

\begin{quotation}
\noindent That the common centre of gravity of the earth, the sun, and all
the planets is immovable.

\noindent For (by corollary iv of the Laws) that centre either is at rest,
or moves uniformly forwards in a right line; but if that centre moved, the
centre of the world would move also, against the Hypothesis.
\end{quotation}

\noindent From the point of view of Galilean space-time, having argued, by
Corollary IV, that the centre of mass of the solar system moves inertially, Newton
took it as a matter of \textit{convention} that it be considered at rest. I
shall come back to this use of Corollary IV shortly.

Newton's appeal to absolute space was excoriated by Mach in his early
writings on mechanics. It was considered by Hans Reichenbach and other
members of the Vienna circle as purely metaphysical (one might have thought
logical empiricists would look on Newton's conventionalist maneuvering with
more sympathy). But philosophers of science have been more forgiving of
Newton since: Howard Stein, in particular, did much to rehabilitate him.

How does Galilean space-time stand up to Leibniz's critique? A general
approach to the treatment of exact symmetries is this: insist that
physically real quantities, and by extension, physically real properties and
relations built out of those quantities, as specified by a physical theory,
be\textit{\ invariant} under the symmetries of that theory.\footnote{%
I have defended this reading of exact symmetries in physics in several
places; see, e.g. Saunders [2003a,b], [2006], [2013].} In the Galilean case
that lets in relative angles and distances, (magnitudes of) relative
velocities, and (magnitudes of) absolute accelerations; it disallows
absolute directions and absolute speeds. The symmetry group of the
space-time is bigger, and now includes the symmetries of the relativity
principle; and the theory (in terms of invariants) no longer speaks of
absolute positions or velocities, but only of (magnitudes of) absolute
accelerations. We then have all the quantities ordinarily thought to be
measurable in NTM, and get rid of the ones forever underdetermined by any
possible observation.

Commitment to invariant quantities does not in itself rule on the reality of
points of space-time; for example, such points can still be relationally
discerned by invariant quantities (again:\ relative distances, angles, and
intervals in time). But if space-time points are to bear spatiotemporal
relations to material particles, as substantivalists assume, the difficulty
recurs:\ such relations will be changed by symmetry transformations,
actively construed -- by relative drags between material bodies and points
of space-time -- hence they are not physically real. The same difficulty
arises for Minkowski space and for general relativistic space-times with
symmetries.\footnote{%
I do not myself think this difficulty is insurmountable, however, for
reasons sketched in my [2003a,b]. See my [2013 Section 3.2] for an analogous
difficulty in the case of permutation symmetry.}

So much is familiar ground. We are about to move on to new territory, and to
the question of what Newton's laws really allow us to determine. But to that
end we will need some of the machinery of affine-space theory, and it will
be helpful to first set it up for the familiar space-times just encountered.%
\footnote{%
We could use the more usual perspective of a differentiable manifold, but
besides keeping the mathematics as simple as possible, I am concerned to use
methods as close as possible as those of \textit{Principia} (an
`affine-space plus' theory in Stachel's [1993] terminology).}

An affine space has much of the structure of a vector space. It is a set in which pairs of points define vectors, used in turn to define the straight lines talked of in the Laws -- the straight line
motions, real or imagined, of bodies moving inertially. Quite generally, let 
$\mathcal{X}$ be a set, let $\mathbb{V}^{n}$ be an $\mathit{n}$-dimensional
vector space, and let $+$ on $\mathcal{X}$ be an action of $\mathbb{V}^{n}$
on $\mathcal{X}$ which is:

\begin{itemize}
\item transitive, that is, for any $a,b\in \mathcal{X}$, there is a vector $%
\overrightarrow{c}$ $\in \mathbb{V}^{n}$ such that $a+\overrightarrow{c}=b$.
\item free, that is, if for some $a\in \mathcal{X}$, $a+\overrightarrow{c}=a+%
\overrightarrow{d},$ then $\overrightarrow{c}=\overrightarrow{d}.$
\end{itemize}
\noindent If free, the vector in $\mathbb{V}^{n}$ (which exists by
transitivity) is unique,\ denote $a-b$.

The structure $\mathcal{\langle X},\mathbb{V}^{n},+\rangle $ is an \textit{%
(n-dimensional) affine space}, denote $\mathcal{A}^{n}$. A notion of
parallelism for pairs of points can now be defined: $a,b$ is\textit{\
parallel} to $a^{\prime },b^{\prime }$' if $a-b=$ $a^{\prime }-b^{\prime }$.
A transformation $g:$ $\mathcal{A}^{n}\rightarrow \mathcal{A}^{n}$ is\textit{%
\ affine} if it preserves parallels, i.e. if $a-b=a^{\prime }-b^{\prime }$,
then $g(a)-g(b)=g(a^{\prime })-g(b^{\prime })$.

If $\mathbb{V}^{n}$ is equipped with the Euclidean scalar product with norm $%
\left\vert .\right\vert $, then a relative distance function $h$ is defined
on $\mathcal{A}^{n}\times $ $\mathcal{A}^{n}$ by $h(a,b)=\left\vert
a-b\right\vert _{n}$. Since $\mathcal{A}^{n}$ is affine, $h$ is
non-degenerate and satisfies the triangle inequality: it is a \textit{metric}
on $\mathcal{A}^{n}$. The resulting space $\langle \mathcal{X},\mathbb{V}%
^{n},+,h\rangle $ is \textit{n-dimensional Euclidean space.}

It is now clear how to proceed. Space-time should be a $4-$dimensional
affine space (to define straight-line inertial motions), but if it is to
respect the relativity principle its Euclidean structure should be limited
to simultaneous $3-$dimensional surfaces with no preferred velocity (so no
decomposition of vectors in $\mathbb{V}^{4}$ into a spatial and temporal
part). Define \textit{neo-Newtonian} space-time as the structure $\langle 
\mathcal{X},\mathbb{V}^{4},+,\mathbb{W},h,t\rangle $, where

\begin{itemize}
\item $\mathbb{V}^{4}$ is a $4$-dimensional real vector space
\item $+$ is a free, transitive action of $\mathbb{V}^{4}$ on $\mathcal{X}$
\item $\mathbb{W}$ is a $3$-dimensional subspace of $\mathbb{V}^{4}$
\item $h$ is a Euclidean inner product on $\mathbb{W}$, defining a norm $%
|.|_{3}$
\item $t$ is a Euclidean inner product on the quotient space $\mathbb{V}%
^{4}/W$, defining a norm $|.|_{t}$.
\end{itemize}

\noindent Since it is a $4$-dimensional affine space the notion of straight
line, and straight-line parallelism, is well-defined. The lines which (as
vectors) lie in $W\ $connect simultaneous points in $\mathcal{X}$, i.e.
points $a$, $b$ such that $a-b\in W$; and for such points $h$ defines a
metric $h(a,b)=\left\vert a-b\right\vert _{n}$ on points in $\mathcal{X}$.
We could, indeed, have begun with a simultaneity relation on $\mathcal{X}$
and defined $W$ in its terms.\footnote{%
The method followed by Stein [1967].}

The affine transformations that preserve $S,t$ and $h$ are the\textit{\
Galilean transformations} (in which I include mirroring and time reversal).
Defining a rectilinear coordinate system $\overrightarrow{x},t$ on $\mathcal{%
X}$ so as to assign equal times to simultaneous points, they are:%
\begin{eqnarray}
\overrightarrow{x} &\rightarrow &R\cdot \overrightarrow{x}+\overrightarrow{u}%
t+\overrightarrow{d} \\
t &\rightarrow &t+s,t\rightarrow -t  \notag
\end{eqnarray}

\noindent where $R$ is a matrix of determinant $\pm 1$, $\overrightarrow{u}$, 
$\overrightarrow{d}$ are $3$-dimensional Euclidean vectors, and $s$ is a
real number. Excluding the inversions, these transformations are
parameterized by 10 real numbers, the (proper, orthochronous, inhomogeneous)
Galilean group. In this way we turn our 'affine-space plus' into a
differentiable manifold (Galilean space-time). The measurable quantities --
the invariants of these transformations -- are time intervals, instantaneous
relative distances and their derivatives, and (norms of) relative velocities
and absolute accelerations, and derivatives.

What then is Newtonian space-time? It is the structure $\langle \mathcal{X},%
\mathbb{V}^{3}\oplus \mathbb{V}^{1},+,h,t\rangle $; that is, it is
neo-Newtonian space-time with the additional structure that any vector in $%
\mathbb{V}^{4}$ is uniquely decomposed into the sum of a vector in $\mathbb{V%
}^{3}$ (with the Euclidean metric $h$) and a vector in $\mathbb{V}^{1}$
(with the temporal metric $t$). Then any two points $a$,$b$ in $\mathcal{X}$
have a unique spatial separation, namely $h(a,b)=|P(a-b)|_{3}$, where $P$ is
the projector from $\mathbb{V}^{3}\oplus \mathbb{V}^{1}$ to $\mathbb{V}^{3}$.

Invariance of the decomposition $\mathbb{V}^{4}$=$\mathbb{V}^{3}\oplus 
\mathbb{V}^{1}$ narrows the space-time symmetries to the 7-dimensional
subgroup consisting of rotations in space and translations in space and
time. The relativity principle may be a symmetry of the laws, but it is not
a symmetry of the space-time. Only with neo-Newtonian (or Galilean)
space-time are the two correctly aligned. Or are they? \bigskip \noindent \bigskip

{\Large 3. An inconvenient truth\smallskip \bigskip }\footnote{%
The first part of this section draws heavily on Barbour [1989 Ch.12, 1999].}

\noindent By the late 19th century, analytical methods took centre stage,
and increasingly group theory. The concept of inertial coordinates was the
crucial one, and of inertial frame. The latter term was first coined by
Ludwig Lange in 1885, taking up a suggestion by Carl Neumann in his
habilitation address of 1869. Neumann's work, published as `On the
principles of the Galilean-Newtonian theory' in 1870, began the debates over
the foundations of NTM by Mach and others that cleared the way for
Einstein's discoveries.

Neumann's monograph was followed soon after by Ernst Mach's \textit{History
and Root of the Principle of the Conservation of Energy}; both argued for
the need for operational definitions in mechanics. The following year the
mathematician Peter Tait offered the following construction (Tait [1983]):
let there be three force-free bodies, no two of which are at relative rest,
that are non-collinear; let the sequence of equal increments in the relative
distance of any two serve as a clock; then their relative distances and
angles with the third, if inertial, must satisfy certain constraints over
time. In fact the first three instantaneous relative configurations --
`snapshots' -- are unconstrained (any will do, inertial or no); only with
the fourth is there a constraint, amounting to a test of whether the three
bodies are really inertial.

Tait's construction was followed by several others, among them one by Lange.
But its weakness as an operational definition of inertial motion was also
clear, for it depended on the availability of force-free (`fundamental')
bodies. Which, precisely, were these? Unlike every other force,
gravitational force cannot be screened off. The proposal, like Newton's
appeal to absolute space, was a purely conceptual device: it had nothing to
do with anything that could in practise be measured. As Bertrand Russell
complained:

\begin{quotation}
\noindent If motion means motion relative to fundamental bodies (and if not,
their introduction is no gain from a logical point of view), then the law of
gravitation becomes strictly meaningless if taken to be universal -- a view
which seems impossible to defend. The theory requires that there should be
matter not subject to any forces, and this is denied by the law of
gravitation. (Russell [1903].)
\end{quotation}

The situation is different if we use Newton's methods. Rather than
force-free, he used gravitating bodies, principally the earth, sun and moon.
Because of the complexity of this system he made use of an iterative
procedure, but in essence the method was an operational one. Thus to a first
approximation he used the area law (Kepler's second law), itself derived
from his principles for two bodies subject only to centripetal forces, to
define a clock (with equal areas swept out in equal times). Time defined in
this way, taking into account perturbations introduced by a third (and
further bodies, notably jupiter), is \textit{ephemeris time}. It is time
defined so as to make the observed relative angles subtended by the planets
at the earth satisfy Newton's equations of motion. In the early 1950s it had
acquired sufficient precision to replace sidereal time as the standard of
time. It was replaced in turn by atomic clock time in 1972.

Does that secure the status of Proposition XI, but replacing `at rest' by
`inertial'? No, for the same problem returns: the solar system as a whole is
no more free from gravitational forces than the `fundamental bodies' of
which Russell spoke. The sun and planets are acted on by bodies beyond the
solar system: in principle, arbitrarily many.\footnote{%
Recognition of this point was slow. The question of the proper motions of
stars among themselves was widely discussed by the mid 19th century, as we
learn from Mary Somerville's \textit{On the Connexion of the Physical
Sciences} [1840] - but not much before, and not even then in terms of their
action on the solar system. (I am grateful to George Smith on this point).
Nor was the universality of gravity taken for granted: Somerville cited the
motions of binary stars as conforming to Newton's laws as evidence that the
stars, and not just bodies in the solar system, truly gravitate.} Newton's
reasoning in deriving Proposition XI, even granting Hypothesis I, was fallacious. Here is
Corollary IV to the Laws, on which the proof turns:
\begin{center}
COROLLARY IV
\end{center}

\begin{quotation}
\noindent The common centre of gravity of two or more bodies does not alter
its state of motion or rest by the actions of the bodies among themselves;
and therefore the common centre of gravity of all bodies acting upon each
other (excluding external actions and impediments) is either at rest, or
moves uniformly in a right line.
\end{quotation}

\noindent The proviso `excluding external actions' robs it of any
application to local gravitating bodies, like the planets and sun, which
cannot be screened from the rest of the universe.

Are the accelerations thus produced at least likely to be small, so that
Proposition XI (given Hypothesis I) holds to a good approximation? The acceleration due to
the stars in our local galaxy is certainly tiny -- no more than $10^{-10}$
msec$^{-2}$ -- but include ever-larger regions of space filled with stars
and their influence builds. It is the same as with Olbers' paradox: the
influence of remote stars falls off as the inverse square, but the number of
stars at a given distance, if uniform, increases as the square. The result
is an acceleration that scales linearly with the diameter of the universe,
given an approximately uniform mass distribution.

Any elementary observation that escaped Newton should be spelt out in full.%
\footnote{%
It was also missed by Stein. Having correctly observed `Newton's theory affords a
way of assigning kinematical states up to a Galilean transformation, on
condition that one has succeeded in accounting completely for the relative
motions by a system of action-reaction pairs, and on the further condition
that there is no reason to suspect the system in question to be subject to
an outside influence imparting equal accelerations to all its members',
Stein [1977] continued: `Newton had the good luck to find such a system:
namely, the solar system; and the skill to effect its thorough dynamical
explication.' It would be by good luck indeed were this to be true.}
Suppose, for simplicity, matter is uniformly distributed with density $\rho $
over an enormously large sphere, and suppose the solar system is located
within the sphere at distance $R$ from its centre. The gravitational forces
due to the matter at distance greater than $R$ cancels, leaving only that of
the matter in the interior. This mass, denote $M$, is as a function of $R$:%
\begin{equation*}
M(R)=\frac{4}{3}\pi R^{3}\rho .
\end{equation*}%
The gravitational action of the sphere of mass $M$ on the solar system (by
hypothesis located on the boundary of the sphere) is the same as if all the
mass were concentrated at the centre, and hence produces an acceleration $a$
(using the proportionality of gravitational to inertial mass):%
\begin{equation}
a=\frac{M(R)G}{R^{2}}=\frac{4}{3}\pi G\rho R.
\end{equation}%
The result is not materially effected on removing the simplifying symmetries
(so long as the mass density is roughly uniform). For sufficiently large $R$%
, the acceleration of the solar system will as large as you like -- or do 
\textit{not} like.

But there is always the centre of mass frame for the universe as a whole,
however large, so long as the total mass is finite. To this Corollary IV applies
exactly and without ambiguity. Relative to this frame, whatever it is, if
stellar densities are similar to those of our local Hubble volume, the
acceleration of the earth due to the rest of the universe swamps that due to
the sun for a universe of size $10^{38}$ m and greater. As it happens, that
is much larger than the \textit{visible} universe (by about twelve orders of
magnitude) -- but there is every reason to think the actual universe is much
larger, if not in fact infinite.

What if the actual universe\textit{\ is} infinite?\ In that case there is a
new difficulty: not so much that the `true' acceleration of the solar system is enormously large and unknown, but that it is undefined. This problem was first recognized by Hugo Seelinger in 1894, and was much discussed subsequently. By the above method the acceleration of the
solar system due to the rest of the universe can be given any number
whatsoever.\footnote{%
This was the inconsistency discussed by Norton [1993], [1999].} The problem was considered so severe that some concluded the
inverse square law itself should be modified. Shortly
after, Einstein made the same suggestion, but for a very different reason.
He too assumed the universe was approximately homogeneous, but in terms of
his newly discovered theory of GTR; and in that context he saw it could not
possibly be static. The solution was to modify the field equations, thus
introducing his `cosmological constant'. But now exactly the same reasoning
applied to Newton's theory of gravity, which Einstein used as a warming-up
exercise. The modification to Newton's theory of gravity eliminated the
divergence of Eq.(6)\ as $R\rightarrow \infty $.\footnote{%
Eq.(2) is replaced by: 
\begin{equation*}
\nabla ^{2}\phi +\Lambda \phi =-4\pi G\rho .
\end{equation*}%
Here $\Lambda $ is the cosmological constant.}

Pursuit of these questions led directly to big-bang cosmologies on the one
hand (because Einstein's solution was ineffectual), and to Newton-Cartan
theory on the other (because Einstein's geometric methods were so
successful). We saw in section 1 how the problem of acceleration is resolved in
that setting. But let us not take our eye off the ball. There is not and never
was any evidence to show the universe is less than $10^{38}$ m in size. Even
in the favourable case, supposing we live in an island universe, by Newton's
principles we never had good reason to think the solar system or any other
concretely-defined frame of reference was even approximately inertial. That
is, the central concept of NTM, the concept of an inertial frame of
reference, was never in practice operationally defined by Newton's
principles, not even approximately.
\bigskip
\bigskip

{\Large 4. Corollary VI\smallskip }$\bigskip \nolinebreak $

\noindent The conclusion just re\nolinebreak ached has an air of paradox. It
seems self-defeating if true: how is it NTM \textit{was}, in fact,
successfully applied?

In only one place in \textit{Principia} did Newton raise the question of the
influence of the stars on the solar system: in Corollary II to Proposition IV of Book 3.
The proposition itself was that the aphelions and nodes of the orbits of the
planets are fixed (the aphelion of an orbit is its furthest point from the
sun; its nodes are the points of intersection of the plane of an orbit with
the ecliptic). The result was proved in Book 1 (at Proposition XI) in the
approximation in which the mutual actions of the planets and comets among
themselves are neglected, but Newton stated it again in Book 3 to draw two
corollaries. The first is that `the fixed stars are immovable, seeing they
keep the same position to the aphelions and nodes of the planets' --
meaning, the fixed stars are non-rotating with respect to the aphelions and
nodes, and therefore absolutely non-rotating.\footnote{%
Newton had, however, in Phenomenon I, II and IV, assumed the stars to be at
rest. HIs argument from induction for his theory of gravity was a complex
one. See DiSalle [2006 Ch.2] and , for the most comprehensive study to date,
Harper [2012].} The second reads:

\begin{quotation}
\noindent And since these stars are liable to no sensible parallax from the
annual motion of the earth, they can have no force, because of their immense
distance, to produce any sensible effect in our system. Not to mention that
the fixed stars, everywhere promiscuously dispersed in the heavens, by their
contrary attractions destroy their mutual actions, by Prop.lxx, Book 1.
\end{quotation}

\noindent On the first point, we must infer Newton never made the
calculation Eq.(6); he simply failed to see that the fall-off as the inverse
square of the gravitational force with distance is cancelled by the
quadratic increase in the number of stars. But on the second the error is
more subtle. Proposition LXX of Book 1 proved that the gravitational force
vanished everywhere in the interior of a uniform shell of matter; Newton's
reasoning, then, would seem to be that the stars are approximately uniformly
distributed (`promiscuously dispersed'), so their action may be considered
as that due to a concentric system of such shells with the sun at their
centre proceeding to infinity. Given all this, their influence \ is shown to
be zero.

We used Proposition LXX ourselves, in the case of an island universe, but in that
case the assumption that the solar system is at the centre of the universe
is clearly unwarranted. In the infinite case the assumption does not even
make sense. Since homogeneous and infinite, no point is at the centre. If
one applies the same construction anyway, one can build a system of spheres
about any point whatsoever, so points at completely arbitrary distances from
the solar system. Whereupon by the argument of Eq.(6) the acceleration can
be anything and we are back to Seelinger's problem.

We are also back to Einstein's problem. Such a universe cannot possibly be
static. Newton recognized rather better than did Einstein that here a
balancing act was impossible -- that nothing short of divine providence
could ensure a static universe.\footnote{%
Newton wrote:'That there should be a central particle, so accurately placed in
the middle as to be always equally attracted on all sides, and thereby
continue without motion, seems to me a supposition fully as hard as to make
the sharpest needle stand upright on its point upon a looking glass. For if
the very mathematical centre of the central particle be not accurately in
the very mathematical centre of the attractive power of the whole mass, the
particle will not be attracted equally on all sides. And much harder is it
to suppose all the particles in an infinite space should be so accurately
poised one among another, as to stand still in a perfect equilibrium. For I
reckon this as hard as to make, not one needle only, but an infinite number
of them (so many as there are particles in infinite space) stand accurately
poised upon their points. Yet I grant it possible, at least by a divine
power; and if they were once to be placed, I agree with you that they would
continue in that posture without motion forever, unless put into new motion
by the same power.' (Letter Jan. 17, Bentley [1838 p.208].) Einstein's solution (by means of a cosmological constant) required just such a balancing act; small wonder he later called it `the greatest
blunder of my life' (on this point see Bianchi and Roveli [2010]).} That
sounds like a reductio, but it is clear from correspondence with Bishop
Bentley that Newton welcomed the conclusion. Like Bentley, Newton looked to
science for arguments for the existence of God.

But our concern is with the definition of inertial frames, and on this point
Newton had more to say. The remark is to be found not in the \textit{%
Principia} itself but in an informal draft of Book 3, published posthumously
in English translation as \textit{The System of the World }in 1728:

\begin{quotation}
\noindent It may be alleged that the sun and planets are impelled by some
other force equally and in the direction of parallel lines, but by such a
force (by Cor.vi of the Laws of Motion) no change would happen in the
situation of the planets one to another, nor any sensible effect follow: but
our business is with the causes of sensible effects. Let us, therefore,
neglect every such force as imaginary and precarious, and of no use in the
phenomena of the heavens\ldots (Cajori [1938 Sec.8].)
\end{quotation}

\noindent Corollary VI to the Laws is the statement:

\begin{center}
COROLLARY VI
\end{center}

\begin{quotation}
\noindent If bodies moved in any manner among themselves, are urged in the
direction of parallel lines by equal accelerative forces\footnote{%
Meaning, forces that produce equal accelerations (see Section 5).}, they
will all continue to move among themselves, after the same manner as if thy
had not been urged by those forces.
\end{quotation}

\noindent The proof was brief and there was no comment on it in the Scholium
that followed. It was used to prove Proposition III of section II (and Proposition LVII
and LXIV in later sections of Book 1); no other reference to it is to be
found in \textit{Principia}. But Proposition IIIi was important to the argument for
the inverse square law of force for gravity in Book 3. The proposition
is:

\begin{center}
PROPOSITION III
\end{center}

\begin{quotation}
\noindent Every body [A], that by a radius drawn to the centre of another
body [B], however moved, describes areas about that centre proportional to
the time, is urged by a force compounded of the centripetal force tending to
that other body, and of all the accelerative force by which that other body
is impelled.
\end{quotation}

\noindent That is, given the premise, the force by which body A is moved is
compounded of a centripetal force, and of whatever force is needed to make
it accelerate at the same rate as is B. But according to the Phenomena of
Book 3 the moons of jupiter \textit{do }describe equal areas in equal times,
so they must have those forces acting on them, so in particular (apart from
the force that makes them accelerate at the same rate as jupiter) the moons
experience a centripetal force towards jupiter. This is the first part of
Newton's deduction from phenomena of the force of gravity as centripetal and
falling off as the inverse square.

But how is it that all the moons of jupiter are accelerated the
same as jupiter by the sun? The answer, of course, lies in Galileo's
equivalence principle -- the principle that in the absence of friction, all bodies
are accelerated the same by a uniform gravitational field, whatever their
composition. In Newton's terms that meant the constancy of the ratio of
gravitational mass $\mu $ to inertial mass $m$; and it is to this that
Newton immediately turned after establishing the force law of gravity, at
Proposition VI of Book 3.\footnote{%
Although he had intimated it much earlier, at Definition VII of Book 1.} Newton
argued for the universality of this ratio at length, on purely empirical
grounds, with reference to experiments with pendulums that he had himself
conducted.\footnote{%
An argument from induction, in Newton's sense, meeting the requirements of
his Rules of Reasoning in Philosophy (in particular the third).}

If, instead, the gravitational force is taken as given, the key to applying
the Laws to the jupiter system is not so much Corollary VI but the combination of
Corollary VI with Galileo's equivalence principle, which may be stated as:

\begin{center}
COROLLARY VI*
\end{center}

\begin{quotation}
\noindent If bodies moved in any manner among themselves, are urged in the
direction of parallel lines by equal gravitational forces due to outside
bodies, they will all continue to move among themselves, after the same
manner as if they had not all been urged by that force.
\end{quotation}

\noindent In this by `equal gravitational force' I mean equal force per unit
gravitational mass, denote $\overrightarrow{g}$, so that acting on a body of
inertial mass $m$ and gravitational mass $\mu $ the force is $\mu 
\overrightarrow{g}$. The resulting acceleration is 
\begin{equation*}
\frac{d^{2}\overrightarrow{x}}{dt^{2}}=\frac{\mu }{m}\overrightarrow{g}.
\end{equation*}

\noindent From the constancy of $\mu /m$ for any material body, whatever its
constitution, and given that $\overrightarrow{g}$ is uniform, the condition
of Corollary VI  is met. Since the gravitational force of the sun is approximately
uniform over the dimensions of the jupiter system, we can conclude that the
motions of jupiter and its moons `among themselves' will be the same as if
the centre of mass system were moving inertially. By Corollary V, these are the
same as if the centre of mass were at rest, whereupon Newton's laws apply.

It needs only a small rearrangement of this reasoning to conclude:

\begin{center}
EQUIVALENCE PRINCIPLE
\end{center}

\begin{quotation}
\noindent If bodies moved in any manner among themselves are described in
relation to an accelerating but non-rotating frame of reference, they will
all move in relation to that frame as if acted on by uniform gravitational
forces, producing the opposite acceleration.
\end{quotation}

\noindent With that we have a version of Einstein's equivalence principle --
and the germ of an explanation for the universal proportionality of
gravitational to inertial mass. However, we are now remote not so much from
Newtonian concepts, as from his purpose in Book 3 of \textit{Principia} --
which was to \textit{deduce} the inverse square law from the observed
motions.

Corollary VI in conjunction with Newton's dynamical determinations of a state of
non-rotation and of equal time intervals\footnote{%
At Proposition 1 Book 1, deriving Kepler's second law (see Barbour [1989 pp.546-56]
for an illuminating discussion).} solves the problem of inertial frames for
gravitational physics. The reason that NTM\ works in application to the
solar system, without ever giving an operational meaning to the notion of
inertial frame, is that none is needed: all that matters is that motions are
referred to a freely-falling non-rotating frame. That, to an excellent
approximation, can be defined in terms of the dynamical behaviour of the
solar system -- using Proposition XIV of Book 3 -- regardless of the distribution
of matter in the rest of the universe. All that is needed is that the stars
and galaxies are sufficiently distant so as to give rise to gravitational
fields approximately uniform over the dimensions of the planetary orbits;
whether or not they are changing in time -- and, indeed, whether or not the
universe is infinite -- is irrelevant.

But now notice Corollary VI* together with Corollary VI\textit{\ also} solves the
problem of inertial frames for non-gravitational forces as well -- so long
as they are described by Newton's laws. The difference, on going over to
non-gravitational physics, is that there is no analog of the constancy of
gravitational to inertial mass; Corollary VI* cannot be used to justify the
neglect of long-range forces other than gravity. But precisely because
charge-to-mass ratios of every other force\textit{\ do} vary, the existence
of such forces would show up in the relative motions of bodies, whether
referred to an inertial frame or -- and this is the crucial point, by Corollary VI
-- to an accelerating but non-rotating frame. So long as the latter (as a
material system - say, a system of gyroscopes in orbit around the earth) can
be screened off from any non-gravitational forces that might produce a
torque, and so long as Newton's laws apply, a dynamical analysis will yield
a criterion of nonrotation, and thereby define a nonrotating frame. In
principle such coordinates can be extended arbitrarily and used to describe
any bodies, anywhere in the universe.

What happens if the bodies comprising the freely-falling frame \textit{cannot%
} be screened off from nongravitational forces due to distant bodies
producing a torque? If there were such forces then there would be a
difficulty. It may be, in such a world, that nothing short of the centre of
mass frame of the entire universe would do -- supposing it is a finite
universe. In a world like that it may well be that physics as we know it
would not be possible. But that world, thankfully, is not ours. The only
long-range force other than gravity is electromagnetism, which (except in
extreme cases) can always be screened off. And anyway, because charges of opposite sign tend to cancel, and those of the same sign repel, we have good reason to believe there are no large remote sources of this kind.

But if all of this is true, is there really any need of `real' inertial
frames, that are anyway operationally unavailable? Why not make do with
local, freely-falling, non-rotating frames, as we do in GTR? If need be, the
concept could be idealized as the limiting case of such a frame. The
idealization is safe, as in practise such frames can be concretely realized
(gyroscopes in orbit) to enormous accuracy.\footnote{%
The Gravity B Probe, recently used to test for dragging of inertial frames -
hence as a test of GTR - is an example.} It is also safe even if the univers  is infinite: the further the mass distribution, the more accurately its gravitational influence on the solar systtem is approximated by a uniform gravitational field. Newton's Principia, on our reading, conforms to the maxim that theories be applicable whether or not the universe is infinite. 
\footnote{%
Note added Sep 2016. This is an optimistic reading. And insofar as it is right, there is no guarantee of agreement with the non-relativistic limit of GTR: see Wallace [2016].}

That question carries with it another: what has any of this really to do
with gravity? Given that any rigid motion can be decomposed into a linear
acceleration and a rotation, all that really matters is that the frame of
reference be non-rotating. Free-fall, and the appeal to gravitational
physics and screening from all other forces, is simply the easiest way of
materializing a non-rotating frame. Corollary VI is more fundamental than the
equivalence principle.\bigskip \bigskip

{\Large 5. Kicking away the ladder\bigskip \nolinebreak }

\noindent Of course our reasoning up to this point has been based on the
laws (Axioms, or Laws of Motion, Book 1), which suppose, by wide consent, that
motions can be referred to an inertial frame. Presented analytically, as
from the time of Lagrange, it is assumed there exist rectilinear inertial
coordinates $\langle \overrightarrow{x},t\rangle $ and the equations written
in terms of these. For $N$ particles subject to forces $F$ that satisfy the
third law (and that for simplicity are independent of relative velocities),
we have the system of equations:

\begin{equation}
m_{j}\frac{d^{2}\overrightarrow{x_j}}{dt^{2}}=\sum_{k\neq j=1}^{N}F(%
\overrightarrow{x_j}-\overrightarrow{x_k},t);\text{ }j=1,...,N.
\end{equation}

The symmetry group of this system of equations (assuming the $m_{j}$'s
transform as scalars) is the Galilean group (5). It would seem, then, that
inertial coordinates, in the usual Galilean sense, are needed to set up the
theory ab initio. It may be that in practise we an refer all the motions to
local freely-falling frames, and the motions of the particles among
themselves will all be the same as if referred to an inertial frame, but
this inertial frame has to be defined, at least conceptually, to even write
down the equations.

But the key concept to our reading of NTM\ as a relational theory, that of
`motions among themselves', plays a role here as elsewhere. It is most
naturally expressed in terms of \textit{difference} equations for particle
pairs, in terms of the relative distance vectors 
\begin{equation}
\overrightarrow{r_{jk}}=\overrightarrow{x_j}-\overrightarrow{x_k}.
\end{equation}%
\bigskip From the $N$ equations (7), define the $N(N-1)/2$ equations%
\begin{equation}
\frac{d^{2}\overrightarrow{r_{jk}}}{dt^{2}}=\frac{1}{m_{j}}\sum_{l\neq j}F(%
\overrightarrow{r_{jl}},t)-\frac{1}{m_{k}}\sum_{l\neq k}F(\overrightarrow{r_{kl}},t).
\end{equation}%
Of these only $N-1$ are linearly independent. From such a set, together with
the equation
\begin{equation}
\sum_{k}m_{k}\frac{d^{2}\overrightarrow{x_k}}{dt^{2}}=\sum_{k}\sum_{j\neq
k}F(\overrightarrow{r_{jk}},t)=0
\end{equation}%
the $N$ equations (7)\ can be derived, and vice versa. Eq.(10), the equation
for conservation of total momentum, follows from (7) and Newton's third law (that
$F(\overrightarrow{r}_{jk},t)$ is antisymmetric in $j$ and $k$. It is
invariant under Galilean boosts 
\begin{equation*}
\overrightarrow{x}_{k}\rightarrow \overrightarrow{x_k}+\overrightarrow{u}t
\end{equation*}%
(by Corollary V), as are the $N(N-1)/2$ equations (9). But (8) and (9) are invariant
under the much wider class of symmetry transformations:%
\begin{equation}
\overrightarrow{x_k}\rightarrow \overrightarrow{x_k}+\overrightarrow{f}%
(t)
\end{equation}%
where $\overrightarrow{f}$ is any twice-differentiable vector-valued
function of the time.\footnote{%
Eq.(11), together with the second of (5), were called the `Newtonian group'
by Ehlers [1973].} The imposition of Eq.(10) is what distinguishes `true'
inertial frames from our non-rotating frames. If we dispense with (10), and
allow that the system of equations (8), (9) define the entire theory -- say
a \textit{new} theory, \textit{different} from NTM -- the symmetries of the
equations will be those of Corollary VI, not just Corollary V.

There is, however, something odd about this way of putting it. For was Corollary VI not \textit{already} a consequence of the Laws, just as was Corollary V? Why
think Eq.(7) is the correct expression of Newton's laws, rather than (8) and
(9)? The latter are clearly better suited to define the motions of a system
of bodies \textit{among themselves}. If it was a check on the correct
equations for the expression of Newton's laws that they exhibit the
symmetries of the laws (Corollary V), leading to (7), why isn't it a check that
they have the symmetries of Corollary VI, namely (11), leading to (8) and (9) and
not (7)?

The answer, presumably, will turn on the extent to which we can view the 
\textit{Principia} as at bottom a theory of the relative motions of particle
pairs. As a first step one would like to see if the Definitions and Laws can
be rewritten in a way that obviously translates into Eq.(8)\ and (9), rather
than (7), \textit{and} that preserves intact the results and methods of proof used in
the first two books of \textit{Principia}.

There is an obvious way to do this, namely, simply relativize them -- suppose that
'absolute space' refers to any rigid nonrotating frame, in the same way
that in passing from Newtonian to neo-Newtonian space-time one takes
`absolute space' to refer to any inertial frame. No matter which frame is
chosen, the subsequent geometrical constructions (the main proof procedure
used in \textit{Principia}) go through unchanged. 

But that may seem like a cheat. Is it possible, rather, to interpret the laws more directly as concerning the relative motions of particle pairs? I think that it is -- and to make clear how to do this, here are the laws in modified form, keeping as close as possible to Newton's language,\footnote{%
First law: `Every body continues in its state of rest, or of uniform motion
in a right line, unless it is compelled to change that state by forces
impressed upon it'. Second law: `The change of motion is proportional to the
motive force impressed; and is made in the direction of the right line in
which that force is impressed' (Cajori [1934, 12]).} the laws might be restated as:  
\begin{description}
\item[First Law*] Every body continues in its state of relative rest, or
uniform relative motion in a right line, with respect to another, unless
compelled to change that state by a difference in accelerative\footnote{%
Note added 5 Sep 2016. Accelerative, and not motive, as incorrectly stated in [2013] (my thanks to Tim Maudlin on this point).} forces impressed
upon them.

\item[Second Law*] The change in the relative motion of two bodies is
proportional to the difference in accelerative forces impressed upon them, and is
made in the direction of the right line in which the difference in accelerative
forces is impressed.

\item[Third Law] To every action there is always opposed an equal reaction:
or, the mutual actions of two bodies upon each other are always equal, and
directed to contrary parts.
\end{description}

\noindent (The third law is unchanged.)

As stated, these laws clearly support (8) and (9) and of course the antisymmetry of $F$. But now analagous modifications will have to be made in stating some of their consequences, in particular Corollary IV. From the modified laws above, it follows not that the
centre of mass of all bodies acting upon each is in uniform motion,
`excluding external actions and impediments', but that it is at rest or in
uniform motion relative to another (that is free from external actions and
impediments, or for which the relative external force vanishes). However. the
first part of Corollary IV stands unchanged.

But what does `uniform, straight line motion' mean, if inertial motions (in
the usual sense of `inertial') are undefined? Just this: relative
straight-line motions are those for which the relative distances (lengths of
lines joining pairs of bodies at each instant of time) and relative angles
between lines joining instantaneous bodies at various times, change
in a characteristic way. What way exactly? The way they change under the
usual representation of force-free bodies, when moving inertially. 

In any case, whatever its relation to \textit{Principia}, the system of
equations (8) and (9) together with the antisymmetry of $F$ is of independent interest -- especially if, as seems
to be promised, it yields exactly the same relative motions as do (7) and (10). Can
we kick away the ladder?\ Can we dispense with the first equality of
Eq.(10)?\textit{\ Formally} there can be no obstacle: we are already assured
that any solution to the $N$ equations (7) will yield $N(N-1)/2$ difference
vectors $\overrightarrow{r}_{jk}$ satisfying (9). Conversely, for any solution
to the $N(N-1)/2$ equations (9) satisfying (8) (of which only $N-1$ functions $%
\overrightarrow{r}_{jk}$ are linearly independent), choose$\overrightarrow{f}(t)$ so that its second derivative is 
\begin{equation}
\frac{d^{2}\overrightarrow{f}}{dt^{2}}
=-\frac{1}{M}\sum_{k}m_{k}\frac{d^{2}\overrightarrow{x_k}}{dt^{2}}
\end{equation}%
where $M$ is the total mass, and apply the transformation (11). From the antisymmetry of $F$ Eq.(10) is then satisfied, and the coordinates as functions of time then satisfy the N equations (7). 

One can put it like this: the set of difference equations,
Eq.(8) and (9), for antisymmetric $F$, has a gauge freedom --\ an overall
acceleration, constant in space, as an arbitrary function of time.\footnote{%
As a result Noether's first theorem no longer applies, and the total
momentum is not a conserved quantity - as has already been observed.} Up to
this choice of gauge, solutions to these equations are in 1:1 correspondence with solutions to (7).%
 As Newton said: `Let us, therefore, neglect every such force as imaginary and
precarious, and of no use in the phenomena of the heavens'.\bigskip \bigskip

{\Large 6. Is it relationalism? \bigskip }

\noindent I have urged that a key concept of \textit{Principia} is that of
`motions among themselves', and made this out in terms of difference
equations. Is a theory based on Eqs.(8), (9) and the symmetry of $F$ a distinctively relational
theory? It \textit{appears} so. It is certainly a theory in which no
meaningful notion of the velocity or acceleration of a single body is
available; there are only relative distances and velocities of simultaneous
bodies and their time derivatives, of two or more bodies among themselves.

More than this is needed, however, if by 'relationalism' we mean a genuine eliminativism of reference to points of space or space-time. But just that is what appears to be on the table: certainly the laws, as just modified, make no mention of points of space or space-time; they speak only of quantities of motion in terms of relations among bodies. 

The devil, however, lies in the details. Let us examine these quantities more carefully. As in
Galilean-covariant theories, they include rates of change of the norms of
the $\overrightarrow{r}_{jk}$'s, denote $r_{jk}$, the quantities:%
\begin{equation}
\frac{dr_{jk}}{dt}, \frac{d^{2}r_{jk}}{dt^{2}} \end{equation}%
These require comparisons of spatial lengths at different times. Considering
three or more bodies, one also has rates of change of relative angles, as given by (for unit $r_{jk}$, $r_{kl}$
\begin{equation}
cos^{-1}(\overrightarrow{r_{jk}}\cdot\overrightarrow{r_{kl}})\text{,  } \frac{d}{dt}cos^{-1} (\overrightarrow{r_{jk}}\cdot\overrightarrow{r_{kl}})
\end{equation}%
Quantities of the form (13), (14) were always acceptable to relationalists.

The traditional difficulty has lain rather with rotations -- say, two bodies
rotating about their common centre. If the motion is purely rotational, then
all quantities of the form (13), 14) are zero. Nevertheless the bodies have
non-zero relative velocities, and, referred to inertial frames, non-zero
accelerations. How do we describe their motion on our relationalist reading
of \textit{Principia}? The answer is that the notion of relative velocity is
independent of the principle of inertia, so carries over unchanged, and the
acceleration goes over to relative accelerations of the rotating bodies,
with no need to refer it to an inertial frame. The two bodies have a
relative velocity because the direction of the line connecting them changes
in time, and they have a relative acceleration because the direction of the
relative velocity changes in time. The space-time structure needed to make
sense of rotations is that a spatial direction at one time can be compared
with a spatial direction at another time, with no need of the concept of
straight-line motion or departure from straight-line motion.

So much is obvious if we write the derivative as the limit of a difference
equation:

\begin{equation}
\frac{d}{dt}\overrightarrow{r_{jk}}(t)=\underset{\Delta t\rightarrow 0}{\lim 
}\frac{1}{\Delta t}\left[ \overrightarrow{r_{jk}}(t+\Delta t)-%
\overrightarrow{r_{jk}}(t)\right] .
\end{equation}%
The mistake was to think this must be expressed as 
\begin{eqnarray*}
\frac{d}{dt}(\overrightarrow{x_j}-\overrightarrow{x_k})(t) &=&\underset{%
\Delta t\rightarrow 0}{\lim }\frac{1}{\Delta t}\{[\overrightarrow{x_j}(t+\Delta t)-\overrightarrow{x_k}(t+\Delta t)]-[\overrightarrow{x_j}(t)-\overrightarrow{x_k}(t)]\} \\&=& \underset{\Delta t\rightarrow 0}{\lim }\frac{1}{\Delta t}[
\overrightarrow{x_j}(t+\Delta t)-\overrightarrow{x_j}(t)] -
\underset{\Delta t\rightarrow 0}{\lim }\frac{1}{\Delta t}[(
\overrightarrow{x_k}(t+\Delta t)-\overrightarrow{x_k}(t)]
\end{eqnarray*}%
that is, as the difference between two velocities, each of which must be independently defined. Only the relative velocity as given by Eq(15) is physically meaningful, not velocities of which it is the difference.

Of course Machians go further, and reject quantities like (15) as well. The
comparison of spatial directions at different times is not needed, according to
Barbour [1999], no more than comparisons of time-like vectors in space-time. The solution to a dynamical problem in Barbour's terms is fully
determined given two relative configurations of particles at different times,
without information on their relative orientation. Our form of relationalism
is, from a Machian point of view, a half-way house.\footnote{%
It is a halfway house from the point of view of Leibniz's principles as well. Leibniz usually spoke of temporal orderings rather than temporal intervals, but an ordering in time is still a diachronic notion. If configurations of bodies at different times can be compared as to order, why not compare them as to spatial direction as well? There is no reason -- save that, for Leibniz, all relations, temporal and spatial, are ultimately to be reduced to monadic predications (they should be `internal' relations). That is just what is provided (for temporal orderings and intervals, if not spatial relations) by Machianism in Barbour's sense.} 
Machians will press the epistemic question: how are these
angles between particle pairs at different times determined? Clearly not in
the way angles between lines in space normally are -- the two lines cannot
be simultaneously observed. The answer is, as always:
dynamically. The comparison (for a pure rotation) is in the first instance
made between the relative velocities of the two bodies at neighbouring times, divided by the time interval
-- their relative acceleration. \textit{This} quantity is given by a
centripetal force law in terms of masses and relative distances of bodies
alone. Since (for a pure rotation) their relative distance is constant in
time, so is the angular velocity and the norm of their relative velocity at
each time. The rate of change of relative velocity then determines the
relative velocity (and the angle between the lines joining the two bodies at
two times) uniquely. All these determinations (or rather their norms) are
invariant under boosts to arbitrary non-rotating frames, the transformations (11), as also,
for time-independent force-laws, under the transformations (5).

I suggest it was considerations like this that led Huygens to reconsider his earlier
view, in reluctant agreement with Newton, that cases of pure rotation could
only be understood as motions in relation to absolute space. At the end of
his life, he wrote instead:

\begin{quotation}
\noindent Circular motion is relative motion along parallel lines, where the
direction is continually changed and the distance is kept constant through a
bond. Circular motion in one body is the relative motion of the parts, while
the distance remains constant owing to the bond. (Huygens [1888-1950,
Vol.21, p.507].)
\end{quotation}

\noindent Passages like these were quoted by Stein as evidence of the depth
of Huygens' thinking (Stein [1977]), but on this point there is hardly a
consensus:

\begin{quotation}
\noindent \lbrack T]o analyze rotation in terms of an objective, or
absolute, notion of velocity difference rather than objective, or absolute
velocity is to possess exactly the insight Newton lacked, but it is also to
reject the full-blown relational conception of motion, something that was
beyond the ken of Huygens' philosophical dogmas. (Earman [1989 p.71].)%
\footnote{%
Barbour [1989 p.675] is likewise critical of Huygens on this point.}
\end{quotation}

\noindent Earman seems to equate Huygen's insight with the shift from
Newtonian to neo-Newtonian space-time, but the insight is deeper: only a
comparison of velocities at the same time is needed, not at different times.
We read Huygens as dispensing with the principle of inertia. Earman thought
no theory like this was available in Newton's day.

That claim, I have shown, is debatable. And whether or not a theory based on
(8)\ and (9)\ is a `full-blown' relationalism, it is surely a theory in
which:

\begin{quotation}
\noindent True and simple motion of any one whole body can in no way be
conceived -- what it is -- and does not differ from rest of that body.
(Huygens [1888-1950, fragment 8, vol.16].) \bigskip 
\end{quotation}

{\Large 7. Newton-Huygens space-time \bigskip }

\noindent The limiting process used in Sec.6 shows what is needed in terms
of manifold structures:\ a connection as a rule for comparing spatial
directions at different times -- a rule for the parallel transport of 
directions in \textit{space} (compatible with the temporal and spatial
metrics), not in space-time. Such a construction (with the connection
an `absolute' structure) was given by Earman; he called it `Maxwell
space-time'.\footnote{%
Earman [1989 \S 2.3, \S 4.7]. He still thought it inadequate to
relationalism, however, as flouting his relational requirement R1, that
space-time `cannot have structures that support absolute quantities of
motion'. (But whether rotation is absolute or relative is part of what is here in
contention.)} Here I conclude with a construction that goes the other way.
Galilean space-time is the right manifold corresponding to neo-Newtonian
space-time, an affine-space; what does Maxwell space-time correspond to? It
should be a weaker structure -- call it `Newton-Huygens space-time' -- but
be sufficient for the constructions used in \textit{Principia}.

Its outline should already be clear. We are looking for a structure in which
affine notions and not just metrical ones are restricted to surfaces of
constant time. Define Newton-Huygens\textit{\ }space-time as the structure $%
\langle \mathcal{X},\mathbb{V}^{3},\mathbb{V}^{1},+_{3},+_{1}\rangle ,$ where

\begin{itemize}
\item $\mathbb{V}^{3}$and $\mathbb{V}^{1}$ are real vector spaces of
dimensions $3$ and $1$ respectively, with Euclidean inner products $\langle
.,.\rangle _{3}$, $\langle .,.\rangle _{1}$, respectively.
\item $+_{3}$ is a free (but not transitive) action of $\mathbb{V}^{3}$ on $%
\mathcal{X}$, i.e. for any $a\in \mathcal{X}$, $\overrightarrow{b}$ $\in 
\mathbb{V}^{3}$,  $a+_{3}\overrightarrow{b} \in\mathcal{X}$
\item $+_{1}$ is a free and transitive action of $\mathbb{V}^{1}$ on the
cosets of $\mathcal{X}$ under the action of $\mathbb{V}^{3}$.
\end{itemize}

\noindent The action of $\mathbb{V}^{3}$ on $\mathcal{X}$ defines an
equivalence relation on $\mathcal{X}$, `simultaneous with'; the cosets of $%
\mathcal{X}$ under $\mathbb{V}^{3}$ are time-slices. By construction, $+_{3}$
acts transitively on any time-slice. Hence, since $+_{3}$ is free, for any
simultaneous $a,b\in \mathcal{X}$, there is a unique $\overrightarrow{c}\in 
\mathbb{V}^{3}$ such that $a+_{3}\overrightarrow{c}=b$, denoted $b-a\in 
\mathbb{V}^{3}.$ Now consider two pairs of simultaneous points in $\mathcal{X%
}$. If $a$ is simultaneous with $b$, and $a^{\prime }$ is simultaneous with $%
\ b^{\prime }$ (possibly at a different time), the quantity:

\begin{equation*}
\cos \theta =\frac{\langle (a-b),(a^{\prime }-b^{\prime })\rangle _{3}}{%
|a-b|_{3}|a^{\prime }-b^{\prime }|_{3}}
\end{equation*}%
is well-defined. $\theta $ is the angle between the directions in space $a-b$
and $a^{\prime }-b^{\prime }$. But there is no notion of straight line
connecting non-simultaneous points; the principle of inertia does not
apply.

There remain important questions, above all, moving over to a manifold formulation: What is the relation between a theory of gravity ðand other forcesÞ formulated in Maxwell space-time and one based on Newton-Cartan space- time?\footnote{This question was recently addressed (and partially answered) by Weatherall [2015].}
 In the latter there is a notion of parallelism for time-like as well as space-like vectors. That notion, we must conclude, is dispensable, to be derived, if at all, by fixing of gauge.

That takes us, again, to field theory, which is beyond our remit here. But it raises a conceptual worry. It might be thought that insofar as NCT is approximately empirically adequate, as go the motions of the planets and terrestrial gravitating bodies, there is de facto an empirical standard of parallelism for time-like vectors. A theory based on Maxwell space-time will simply be unable to account for this standard.

That argument, however, is too quick. We have already demonstrated the empirical equivalence of a theory based on Newton-Huygens space-time to one based on neo-Newtonian space-time, although the latter too incorporates a standard of parallelism for time-like vectors. That standard, from the perspective of Newton-Huygens space-time, is not an empirical given: it is unneeded surplus structure. We have already shown how to obtain it in the case of a finite universe by a choice of gauge, Eq.(12). Something similar follows from the point of view of Barbour-Bertotti space-time, in which Eq(7) can be derived by a certain parameterization of the allowed curves in relative configuration space. And of course, according to that theory, the standard of parallelism of space-like vectors, with which Newton-Huygens theory is equipped, is likewise surplus structure, unnecessary to the empirical adequacy of the theory (given that the universe has zero total angular momentum and energy).

So much for the actual world. What of possible worlds, and distinctions among them drawn in NCT, invisible to ours? Take possible worlds each with only a single structureless particle. Depending on the connection, there will be infinitely many distinct trajectories, infinitely many distinct worlds of this kind. But in Newton-Huygens terms, as in Barbour-Bertotti theory, there is only one such world -- a trivial one in which there are no meaningful predications of the motion of the particle at all. Only for worlds with two or more particles can distinctions among motions be drawn. From the point of view of the latter theories, the fault lies with introducing a nontrivial connection -- curvature -- without any source, unrelated to the matter distribution. At a deeper level, it is with introducing machinery -- a standard of parallelism for time-like vectors, defined even for a single particle -- that from the point of view of a relationalist conception of particle motions is unintelligible.

 \bigskip \bigskip 

{\Large References\bigskip \bigskip }

\setlength{\parindent}{-0.7cm}Alexander, H. (ed.) [1956], \textit{The Leibniz-Clarke
Correspondence}:\ New York.

\setlength{\parindent}{-0.7cm}Barbour, J. [1989], `\textit{Absolute or Relative Motion? Vol.1:
The discovery of dynamics}. Cambrige University Press.

--- [1999],\textit{\ The End of Time}, Weidenfeld and Nicolson.

\setlength{\parindent}{-0.7cm} Bentley, R. [1838], \textit{Sermons Preached at Boyle's Lecture}:
Francis MacPherson.

\setlength{\parindent}{-0.7cm} Bianchi, E. and C. Rovelli [2010], `Why all these prejudices
against a constant?', available online at https://arxiv.org/abs/1002.3966.

\setlength{\parindent}{-0.7cm} Brown, H. [2005], \textit{Physical Relativity}: Oxford University
Press.

\setlength{\parindent}{-0.7cm} Cajori, F. [1934], \textit{Sir Isaac Newton's Mathematical
Principles of Natural Philosophy and his System of the World}, trans. A.
Motte: University of California Press.

\setlength{\parindent}{-0.7cm} DiSalle, R. [2006], \textit{Understanding Space-Time}: Cambridge
University Press.

\setlength{\parindent}{-0.7cm} Earman, J. [1989], \textit{World Enough and Space-Time}, MIT\
Press.

\setlength{\parindent}{-0.7cm} Ehlers, J. [1973], `Survey of general relativity theory', in 
\textit{Relativity, Astrophysics and Cosmology}, W. Israel (ed.), Reidel.

\setlength{\parindent}{-0.7cm} Friedman, M. [1983],\textit{\ Foundations of Space-Time Theories}:
Princeton University Press.

\setlength{\parindent}{-0.7cm} Harper, W. [2012],\textit{\ Isaac Newton's Scientific Method:
Turning data into evidence about gravity and cosmology}, OUP.

\setlength{\parindent}{-0.7cm} Hood, C. [1970], `A reformulation of Newtonian dynamics', \textit{%
American Journal of Physics} \textbf{38}, 438-442.

\setlength{\parindent}{-0.7cm} Huygens, C. [1885-1950],\textit{\ Oeuvres Compl\`{e}tes}: Soci\'{e}%
t\'{e} Hollandaise des Sciences.

\setlength{\parindent}{-0.7cm} Knox, E. [2013], ` Effective space-time geometry', \textit{Studies in History and Philosophy of Modern Physics} \textbf{44}, 346– 356.

--- (2014), `Newtonian space-time structure in light of the equivalence principle', \textit{The British Journal for the Philosophy of Science} \textbf{65}, 863–880.

\setlength{\parindent}{-0.7cm} Malament, D. [1995], `Is Newtonian cosmology really
inconsistent?', \textit{Philosophy of Science} \textbf{62}, 489-510.

\setlength{\parindent}{-0.7cm} Norton, J. [1993], `A paradox in Newtonian cosmology', in \textit{%
Proceedings of the 1992 Biennial Meeting of the Philosophy of Science
Association, Vol.2}, M. Forbes, D. Hull and K. Okruhlik (eds.), p.412-20.

--- [1995], `The force of Newtonian cosmology: acceleration is relative',%
\textit{\ Philosophy of Science} \textbf{62}, 511-22.

--- [1999], `The cosmological woes of Newtonian gravitation theory', in 
\textit{The Expanding Worlds of General Relativity}, Einstein Studies,
Volume 7, H. Goenner, J. Renn, J. Ritter, T. Sauer (eds.), p.271-323.

\setlength{\parindent}{-0.7cm} Rosen, G. [1972], `Galilean invariance and the general covariance
of nonrelativistic laws', \textit{American Journal of Physics} \textbf{40},
683-687.

\setlength{\parindent}{-0.7cm} Russell, B. [1903], \textit{Principles of Mathematics}: Allen and
Unwin.

\setlength{\parindent}{-0.7cm} Saunders, S. [2003a], `Physics and Leibniz's principles', in 
\textit{Symmetries in Physics: Philosophical Reflections}, K. Brading and E.
Castellani (eds.): Cambridge University Press. Available online at  Available online at http://philsci-archive.pitt.edu/2012/

--- [2003b], `Indiscernibles, General Covariance, and Other Symmetries', in 
\textit{Revisiting the Foundations of Relativistic Physics: Festschrift in
Honour of John Stachel}, A. Ashtekar, D. Howard, J. Renn, S. Sarkar, and A.
Shimony (eds.): Kluwer. Available online at http://philsci-archive.pitt.edu/459/.

--- [2013], `Indistinguishability', in \textit{Handbook of Philosophy of
Physics}, R. Batterman (ed.): Oxford. Available online at http://arxiv.org/abs/1609.05504 

\setlength{\parindent}{-0.7cm} Sommerville, M. [1840], \textit{On the Connexion of the Physical
Sciences}.

\setlength{\parindent}{-0.7cm} Stachel, J. [1993], `The meaning of general covariance', in\textit{%
\ Philosophical Problems of the Internal and External Worlds: Essays on the
philosophy of Adolf Grunbaum}, J. Earman, A. Janis, G. Massey, and N.
Rescher (eds.): Pittsburgh.

\setlength{\parindent}{-0.7cm} Stein, H. [1967], `Newtonian space-time', \textit{Texas Quarterly} 
\textbf{10}, 174-200.

--- [1977], `Some philosophical pre-history of general relativity', in\textit{%
\ Foundations of Space-Time Theories}, Minnesota Studies in Philosophy of
Science, Vol.8, J. Earman, C. Glymore, J. Stachel (eds.): University of
Minnesota Press.

\setlength{\parindent}{-0.7cm} Tait, P. [1883], `Note on reference frames',\textit{\ Proceedings
of the Royal Society of Edinborough}, Session 1983-84, 743-45.

\setlength{\parindent}{-0.7cm} Vickers, P. [2009], `Was Newtonian cosmology really
inconsistent?', \textit{Studies in History and Philosophy of Modern Physics} 
\textbf{40}, 197--208

\setlength{\parindent}{-0.7cm} Wallace, D. [2016], 'More problems for Newtonian cosmology', PhilSci Archive http://philsci-archive.pitt.edu/12036

\setlength{\parindent}{-0.7cm} Weatherall, J. (2015), `Maxwell-Huygens, Newton-Cartan, and Saunders-Knox Space-Times', \textit{Philosophy of Science} \textbf{83}, 82–92.

\end{document}